\documentclass{aastex63}
\usepackage{amsfonts}
\usepackage{subfigure}
\usepackage{graphicx}
\usepackage{amsmath}

\received{August 03, 2020}
\revised{September 15, 2020}
\accepted{September 17,2020}
\submitjournal{ApJ}

\begin{document}

\title{Effects of modified theories of gravity on neutrino pair annihilation energy deposition near neutron stars}


\author{Gaetano Lambiase}
\affiliation{Dipartimento di Fisica ``E.R Caianiello'', Università degli Studi di Salerno, Via Giovanni Paolo II, 132 - 84084 Fisciano (SA), Italy}
\affiliation{Istituto Nazionale di Fisica Nucleare - Gruppo Collegato di Salerno - Sezione di Napoli, Via Giovanni Paolo II, 132 - 84084 Fisciano (SA), Italy}

\author{Leonardo Mastrototaro}
\affiliation{Dipartimento di Fisica ``E.R Caianiello'', Università degli Studi di Salerno, Via Giovanni Paolo II, 132 - 84084 Fisciano (SA), Italy}
\affiliation{Istituto Nazionale di Fisica Nucleare - Gruppo Collegato di Salerno - Sezione di Napoli, Via Giovanni Paolo II, 132 - 84084 Fisciano (SA), Italy}

\begin{abstract}

We study the
neutrino pairs annihilation into electron-positron pairs ($\nu+{\bar \nu}\to e^- + e^+$) near the surface of a neutron star. The analysis is performed in the framework of extended theories of gravity. The latter induce
a modification of
the minimum photon-sphere radius ($R_{ph}$) and the maximum energy deposition rate near to $R_{ph}$, as compared to ones of General Relativity. These results might lead to an efficient mechanism for generating GRBs.
\end{abstract}


\section{Introduction}

General Relativity (GR) is without any doubts the best theory of the gravitational interaction. Although its predictions have been tested to very high precision \citep{Turyshev:2008ur}, there are still open questions which make GR incomplete. The latter arise at short distances and small time scales (black hole and cosmological singularities, respectively), for which any predictability is lost.
On the other hand, the prediction of cosmic microwave background radiation (CMBR)  and the formation of primordial light elements (Big Bang Nucleosynthesis) certainly represent the greatest success of GR. Despite these fundamental results, 
deviations from GR (hence from the Hilbert-Einstein action on which GR is based) are needed, and new ingredients, such as dark matter and dark energy, are required for fitting the present picture of our Universe \citep{riess,riess1,riess2,riess3,riess4,riess5}. In this respect, one may consider, for example, generalization of Einstein's GR where the gravitational action can be a more general function of the higher-order curvature invariants, $\mathcal{L}\sim f(R,R_{\mu\nu}R^{\mu\nu},\Box^k R,\dots)$ \citep{starobinski,starobinski1,Boehmer:2007kx,odi,odi1,odi2,odi3,odi4,odi5,odi6,odi7,capoz1,capoz2,capoz3,Capozziello:2011et,anu,anu1,anu2}.
In the last years, indeed, several {\it alternative} or {\it modified} theories of gravity have been proposed, which allow at the same time to address the shortcomings related to the Cosmological Standard Model. To give an example, higher-order curvature invariants than the simple Ricci scalar $R$ allow getting inflationary behaviour, removing the primordial singularity, as well as to explain the flatness and horizon problems \citep{starobinski,starobinski1} ( for further applications, see Refs. \citep{cosmo,cosmo1,cosmo2,cosmo3,cosmo4,cosmo5,cosmo6,cosmo7,cosmo8,cosmo9,cosmo10,cosmo11,cosmo12,
cosmo13,cosmo14,cosmo15,odi,cosmo17,cosmo18,cosmo19,cosmo20,cosmo21,cosmo22,cosmo23,cosmo24,cosmo25}). This approach, as well as ones related to it, follows from the fact that the high curvature regime requires that curvature invariants are necessary for building up self-consistent effective actions in curved spacetime \citep{birrell,shapiro,barth}.

It is worth to mention that over the past decade some models have been proposed in which deviations from GR occur at the {\it ultraweak-field} regime by means of screening effects \citep{joyce}. One introduces an additional degree of freedom (typically a scalar field) that obeys a non-linear equation driven by the matter density, hence coupled to the environment.
Screening mechanisms play a non-trivial role in what they allow to circumvent Solar system and laboratory tests by suppressing, in a dynamical way, deviations from GR. In particular, the effects of the additional degrees of freedom (the scalar field) are hidden, in high-density regions, by the coupling of the field with matter
while, in low-density regions, they are unsuppressed on cosmological scales. Screening mechanisms studied in literature are the chameleon mechanism \citep{veltman,veltman1}, the symmetron mechanism \citep{symmetron}, and Vainshtein mechanism \citep{Vainshtein}. New tests of the gravitational interaction may, therefore, provide an answer to these fundamental questions \citep{Buoninfante:2019uwo}.


The aim of this paper is to investigate the effects of modified gravity on the neutrino pair annihilation efficiency\footnote{The role of gravity on the neutrino propagation has been studied both in GR \citep{nuGR,nugr1,nugr2,nugr3,nugr4,nugr5,nugr6,nugr7,nugr8,nugr9} and in modified gravity \citep{nuETG,nuetg1,nuetg2}.}. 
In particular, we focus on the process $\nu+\bar{\nu}\rightarrow e^+ + e^-$, which is important for the delay shock mechanism into the Type II Supernova: indeed, at late time, from the hot proto-neutron star, the energy is deposited into the supernova envelope via neutrino pair annihilation and neutrino-lepton scattering. These processes augment the neutrino heating of the envelope generating a successful supernova explosion \citep{Salmonson:1999es}. Moreover, the process is relevant for collapsing neutron stars and for gamma-ray bursts, for which neutrino pairs annihilation has been considered as one for the possible sources.

Calculations performed in the framework of General Relativity show that the efficiency of the processes $\nu\bar{\nu}\rightarrow e^-e^+$, as compared to  the Newtonian case, increases by a factor $\sim 30$ for neutron stars and $\sim4$ in supernovae~\citep{Salmonson:1999es}. This enhancement, however, is not enough for explaining the observed GRBs, to which we are mainly interested in. As we shall see, the computation of the $\nu\bar{\nu}\rightarrow e^-e^+$ efficiency  in generalized theories of gravity may increase by a factor $\sim 3$ needed to explain the observed GRBs. In this respect, results show that modified theories of gravity might play a not trivial role in the context of  high gravity envelopments, such as near a neutron star.

The paper is organized as follows. In Sec. \ref{Geodetics in a generic external gravitational field} and Sec. \ref{Neutrino annihilation} we introduce the formalism needed to study energy deposition rate in a beyond general relativity framework. In Sec.\ref{Charged Galileon}, \ref{Einstein dilaton Gauss Bonnet gravity}, \ref{Brans Dicke theory}, \ref{Eddington-inspired Born-Infeld black hole solution}, \ref{Born–Infeld generalization of Reissner–Nordstrom solution}, \ref{Higher derivative gravity analytical solution} we show our results for Charged Galileon, Einstein dilaton Gauss Bonnet, Brans Dicke, Eddington-inspired Born-Infeld, Born–Infeld generalization of Reissner–Nordstrom solution and Higher derivative gravity respectively. In Section \ref{Gamma Ray Burst analysis} we discuss the previous results in the framework of gamma-ray bursts. Finally, in Sec.~\ref{Conclusions} we state our conclusions on the phenomena.

\section{Geodetics in a generic external gravitational field}
\label{Geodetics in a generic external gravitational field}

In this Section we express a general way to treat the energy deposition problem. We consider a generic metric of the form
\begin{equation}
g_{\mu\nu}= \left(\begin{matrix} g_{00}& 0& 0& &g_{03}\\0& g_{11}& 0& &0\\0& 0& g_{22}& &0\\0& 0& 0& &g_{33}
\end{matrix}\right) \,\ .
\end{equation}
It is possible to define Local Lorentz tetrad as \citep{Prasanna:2001ie}
\begin{equation}
g_{\mu\nu}=V_{\mu}^iV_{\nu}^j\eta_{ij} \,\ ,
\end{equation}
where
\begin{equation}
V_{\mu i}=\left(\begin{matrix}
\sqrt{g_{00}-g^2_{03}/g_{33}}& 0& 0& &g_{03}/\sqrt{g_{33}}\\
0& \sqrt{g_{11}}& 0& &0\\
0& 0& \sqrt{g_{33}}& &0\\
0& 0& 0& &\sqrt{g_{33}}
\end{matrix}\right) \,\ .
\end{equation}
With the above metric, we have a Lagrangian for a close circular orbit ($\theta=\pi/2$)
\begin{equation}
\mathcal{L}=\frac{1}{2}g_{\mu\nu}\dot{x}^{\mu}\dot{x}^{\nu} \,\ ,
\end{equation}
and the generalized momenta read
\begin{align}
p_0&=\frac{\partial \mathcal{L}}{\partial \dot{t}}=g_{00}\dot{t}+\frac{1}{2}g_{03}\dot{\phi}=-E \,\ ,\\
p_1&=\frac{\partial \mathcal{L}}{\partial \dot{r}}=g_{11}\dot{r} \,\ ,\\
p_3&=\frac{\partial \mathcal{L}}{\partial \dot{\phi}}=g_{33}\dot{\phi}+g_{03}\dot{t}=L \,\ ,
\end{align}
where $E$ and $L$ are the energy and momentum of the particle, respectively. Moreover, the Hamiltonian is defined as
\begin{equation}
2\mathcal{H}=-E\dot{t}+L\dot{\phi}+g_{11}\dot{r}^2=\delta_1 \,\ ,
\end{equation}
where $\delta_1=0$ for null geodetics. With the above definitions, one obtains~\citep{Prasanna:2001ie}:
\begin{align}
U^{3}&=\dot{\phi}=E\left(\frac{L}{E}+\frac{1}{2}\frac{g_{03}}{g_{00}}\right)\left(g_{33}-\frac{1}{2}\frac{g_{03}^2}{g_{00}}\right)^{-1} \,\ ; \\
U^0&=\dot{t}=-\frac{E}{g_{00}}\left[1+\frac{1}{2}g_{03}\left(\frac{L}{E}+\frac{1}{2}\frac{g_{03}}{g_{00}}\right)\left(g_{33}-\frac{1}{2}\frac{g_{03}^2}{g_{00}}\right)^{-1}\right] \,\ ; \\
\dot{r}^2&=\frac{E\dot{t}-L\dot{\phi}}{g_{11}} \,\ ,
\label{dr/dt}
\end{align}
where $L/E=b$ is the impact parameter for a massless particle.\\
It is possible to define the angle $\theta_r$, which is the angle between the trajectory and the tangent velocity in terms of local radial and longitudinal velocities~\citep{Prasanna:2001ie}
\begin{equation}
\tan\theta_r=\frac{v^1}{v^3}=\frac{V_r^1v^r}{V_{\phi}^3v^{\phi}+V_t^3}=\frac{dr}{d\phi}\frac{V_r^1}{V_{\phi}^3+V_t^3/v^{\phi}} \,\ ,
\end{equation}
with $v^{\phi}=U^{\phi}/U^t$. These equations can be solved to find a relation between $b$ and $\theta_r$, and using $dr/d\tau$ obtained from Eq.~(\ref{dr/dt}). In general, one would obtain an equation of the kind:
\begin{equation}
b=f(\cot\theta_r) \,\ .
\end{equation}
In a single orbit, $b$ is constant in each point. Thus, for a particle emitted tangentially from the surface ($\theta_R=\pi/2$), we can write formally that~\citep{Prasanna:2001ie,Salmonson:1999es}:
\begin{equation}
\cot\theta_r=f^{-1}\left(f(0)\right) \,\ .
\label{thetar}
\end{equation}
\subsection{Photosphere}
The photosphere is the last stable circular orbit for massless particles. The conditions at the photosphere are of importance when calculating the energy deposition because the neutrino emission properties sensitively depend on the photosphere temperature. In a circular orbit, defined by the condition
\begin{equation}
\dot{r}^2=f(E,L)=0 \,\ ,
\end{equation}
and solving with respect to $E$, one infers~\citep{Khoo:2016xqv}:
\begin{equation}
E^2=V_{\mathrm{eff}} \,\ .
\end{equation}
To find the minimum radius $R_{ph}$, we have to impose that:
\begin{equation}
\frac{\partial V_{\mathrm{eff}}}{\partial r}=0 \,\ .
\end{equation}
In our work we consider neutrinos as massless particles and thus the neutrino-sphere radius (the spherical surface where the stellar material is transparent to neutrinos and from which neutrinos are emitted freely) is larger or equal than the photosphere radius.
\section{Neutrino annihilation}
\label{Neutrino annihilation}
In this Section, we discuss the relativistic calculation of $\nu\bar{\nu}\rightarrow e^+e^-$ energy deposition. Its rate per unit time and unit volume is given in general~\citep{Salmonson:1999es}
\begin{equation}
\dot{q}=\frac{7DG_F^2\pi^3\xi(5)}{2c^5h^6}(kT(r))^9\Theta(r) \,\ ,
\label{qpunto}
\end{equation}
where $G_F$ is the Fermi constant, $D=1\pm4\sin^2\theta_W+8\sin^4\theta_W$, $\sin^2\theta_W=0.23$ and the plus sign is for electron neutrinos and antineutrinos while the minus sign is for muon and tau type. $T(r)$ is the temperature measured by the local observer and $\Theta(r)$ is the angular integration factor. It is possible to write
\begin{equation}
\begin{split}
\Theta(r)&=\int\int(1-\mathbf{\Omega}_{\nu}\cdot\mathbf{\Omega}_{\bar{\nu}})d\mathbf{\Omega}_{\nu}d\mathbf{\Omega}_{\bar{\nu}}\\&=4\pi^2\int_x^1\int_x^1\left[1-2\mu_{\nu}\mu_{\bar{\nu}}+\mu_{\nu}^2\mu_{\bar{\nu}}^2+\frac{1}{2}(1-\mu_{\nu}^2)(1-\mu_{\bar{\nu}}^2)\right]d\mu_{\nu}d\mu_{\bar{\nu}} \,\ ,
\end{split}
\end{equation}
where $\mu=\sin\theta$, $\mathbf{\Omega}=(\mu,\sqrt{1-\mu^2}\cos\phi,\sqrt{1-\mu^2}\sin\phi)$ and $d\mathbf{\Omega}=\cos\theta d\theta d\phi$. The result is:
\begin{equation}
\Theta(r)=\frac{2\pi^3}{3}(1-x)^4(x^2+4x+5) \,\ ,
\end{equation}
where $x=\sin\theta_r$ that can be obtained from Eq.~(\ref{thetar})\\
\subsection{Redshift}
The neutrino temperature varies linearly with redshift and $T(r)$ is related to the neutrino temperature at the neutrinosphere radius $R$ as~\citep{Salmonson:1999es}
\begin{equation}
T(r)=\frac{\sqrt{g_{00}(R)}}{\sqrt{g_{00}(r)}}T(R) \,\ ,
\end{equation}
with $g_{03}=0$. Otherwise the luminosity varies quadratically with redshift
\begin{equation}
L_{\infty}=g_{00}(R)L(R) \,\ ,
\label{Linfinito}
\end{equation}
and, at the neutrinosphere, the luminosity for a single neutrino specie is given by:
\begin{equation}
L(R)=4\pi R^2\frac{7}{4}\frac{ac}{4}T(R)^4 \,\ .
\end{equation}
Combining these equations with Eq.~(\ref{qpunto}), we obtain:
\begin{equation}
\dot{q}=\frac{7DG_F^2\pi^3\xi(5)}{2c^5h^6}k^9\left(\frac{7}{4}\pi ac\right)^{-9/4}L_{\infty}^{9/4}\Theta(r)\left[\frac{\sqrt{g_{00}(R)}}{g_{00}(r)}\right]^{9/2}R^{-9/2} \,\ .
\end{equation}
The total amount of local energy deposited by $\nu\bar{\nu}\rightarrow e^+e^-$ for a single neutrino flavour for a unit time can be defined as:
\begin{equation}
\dot{Q}=\int_R^{\infty} 4\pi r^2dr\sqrt{g_{11}}\dot{q} \,\ ,
\end{equation}
where we integrate $\dot{q}$ from the neutrinosphere radius $R$ to infinity. It is important to state that $R$ depends on the considered situation and can varies from $R_{\mathrm{ph}}$ to infinity: for example, for a Neutron Star it is possible to consider $R=R_{\mathrm{ph}}$, while for a Supernova $R=4$-$5~M$, whit $M$ the core mass.
In the case $g_{00}=g_{11}^{-1}$, one infers
\begin{equation}
\begin{split}
\dot{Q}=&\frac{28DG_F^2\pi^6\xi(5)}{2c^5h^6}\left(\frac{k^4}{7/4\pi ac}\right)^{9/4}DL_{\infty}^{9/4}\left(\frac{g_{00}(R)^{3/2}}{R}\right)^{3/2}\times \\&\times \int_1^{\infty}(x-1)^4(x^2+4x+5)\frac{y^2dy}{(g_{00}(yR))^5} \,\ ,
\label{calcoloQ}
\end{split}
\end{equation}
where $y=r/R$. It is possible to write Eq.~(\ref{calcoloQ}) as:
\begin{equation}
\dot{Q}_{51}=1.09\times 10^{-5}\mathcal{F}\left(\frac{M}{R}\right)DL_{51}^{9/4}R_6^{-3/2} \,\ ,
\label{contoQ}
\end{equation}
where $\dot{Q}_{51}$ and $L_{51}$ are the total energy deposition and luminosity respectively in units of $10^{51}~\mathrm{ergs~s^{-1}}$, $R_6$ is the radius in units of $10~\mathrm{km}$ and
\begin{equation}
\mathcal{F}\left(\frac{M}{R}\right)=3g_{00}(R)^{9/4}\int_1^{\mathrm{R_{ch}}}(x-1)^4(x^2+4x+5)\frac{y^2dy}{g_{00}(yR)^5} \,\ .
\end{equation}
In the Newtonian case, $\mathcal{F}(0)=1$ and, therefore, the ratio $\dot{Q}_{\mathrm{GR}}/\dot{Q}_{\mathrm{Newt}}=\mathcal{F}(M/R)$.
The general form of $\mathcal{F}\left(\frac{M}{R}\right)$ for $g_{00}\neq g_{11}^{-1}$ is given by
\begin{equation}
\mathcal{F}\left(\frac{M}{R}\right)=3g_{00}(R)^{9/4}\int_1^{\mathrm{R_{ch}}}(x-1)^4(x^2+4x+5)\frac{y^2g_{11}(yR)dy}{g_{00}(yR)^{9/2}} \,\ .
\end{equation}

\vspace{0.2in}

\section{Neutrino deposition in modified gravity}

In this section, we explore the neutrino pair annihilation in different models of modified gravity. The studied effect happens in a strong gravitational field, thus we will consider the black hole (BH) solutions for the chosen theories.

\subsection{Charged Galileon}
\label{Charged Galileon}
We investigate the neutrino pair annihilation in the charged Galileon black holes framework, a subclass of Horndeski theories. Besides
nonminimal coupling between scalar and gravity, the above model also inherits an additional gauge field which couples to the scalar sector nonminimally. The action takes the form of~\citep{Mukherjee:2017fqz}
\begin{equation}
\begin{split}
S=&\frac{1}{16\pi}\int d^4x\sqrt{-g}\Bigg[R-\frac{1}{4}F_{\mu\nu}F^{\mu\nu}+\beta G^{\mu\nu}\nabla_{\mu}\psi\nabla_{\nu}\psi-\eta\partial_{\mu}\psi\partial^{\mu}\psi-\\&\frac{\gamma}{2}\left(F_{\mu\sigma}F_{\nu}^{\sigma}-\frac{1}{4}g_{\mu\nu}F^{\alpha\beta}F_{\alpha\beta}\right)\nabla^{\mu}\psi\nabla^{\nu}\psi\Bigg] \,\ ,
\end{split}
\end{equation}
where $\beta\neq 0$ and $\psi$ is the gauge field. Imposing spherical condition one could obtain an exact solution~\citep{Mukherjee:2017fqz}:
\begin{equation}
\begin{split}
ds^2=&-\left(1-\frac{2M}{r}+\frac{\eta r^2}{3\beta}+\frac{\gamma(Q^2+P^2)}{4\beta r^2}\right)dt^2\\
&+\left(1-\frac{2M}{r}+\frac{\eta r^2}{3\beta}+\frac{\gamma(Q^2+P^2)}{4\beta r^2}\right)^{-1}dr^2\\
&+r^2d\Omega^2 \,\ ,
\end{split}
\end{equation}
where $\eta/3\beta=-\Lambda$, with $\Lambda$ cosmological constant. Thus, to be consistent, $\eta<0$ and $\gamma>\beta>0$. It is also possible to define $\gamma(Q^2+P^2)/4\beta=M^2q$, finding that:
\begin{equation}
g_{00}(r)=g_{11}(r)^{-1}=1-\frac{2M}{r}-\frac{\Lambda r^2}{3}+\frac{M^2q}{r^2} \,\ .
\label{metrica1}
\end{equation}
If $\Lambda\neq 0$, this metric is not flat for $r\rightarrow 0$ and this lead to the existence of a cosmological horizon. 
\subsection*{}
\begin{figure}[h!]
\centering
\includegraphics[scale=0.7]{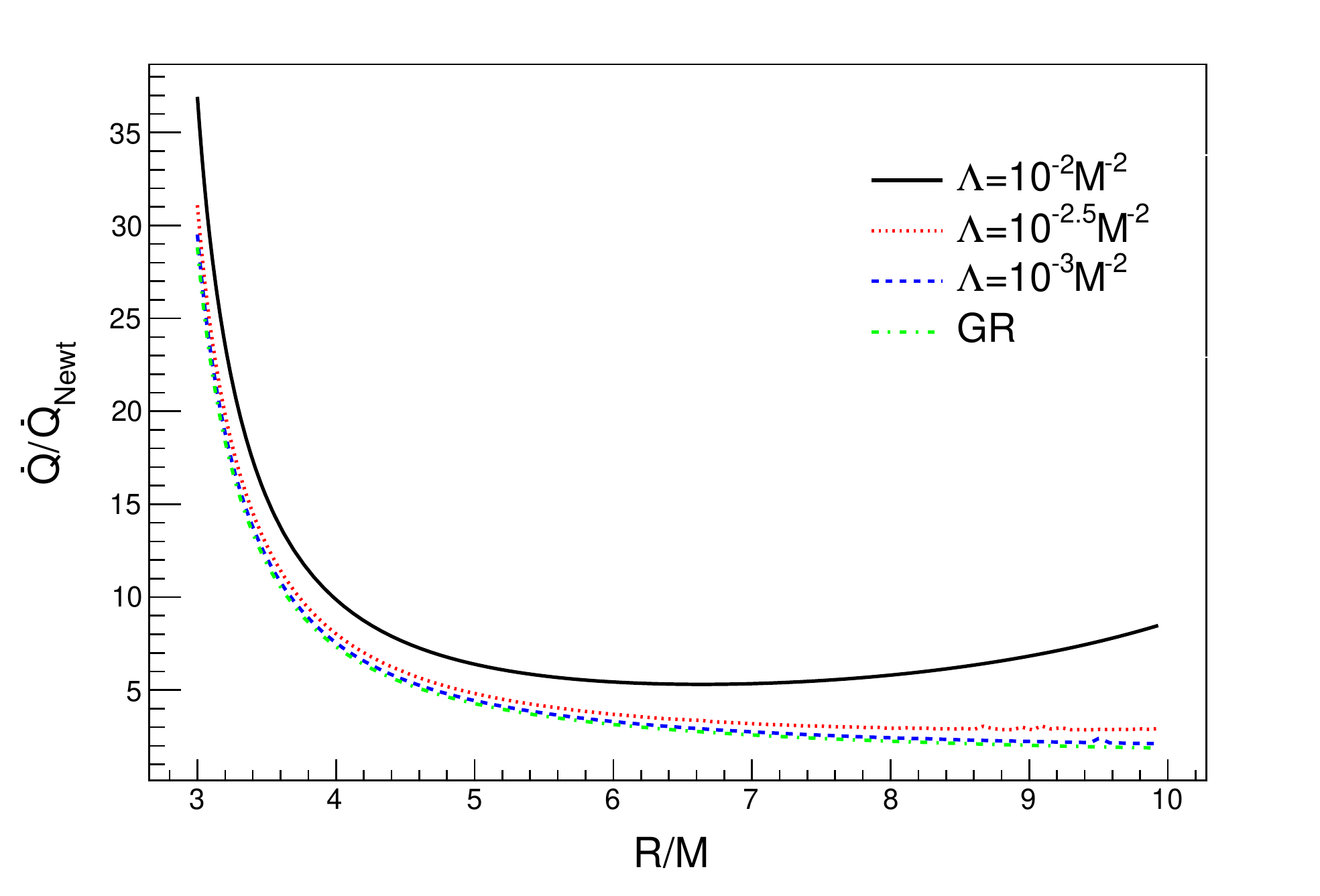}
\caption{Ratio of energy deposition $\dot{Q}$ for metric in Eq.~(\ref{metrica1}) to total Newtonian energy deposition $\dot{Q}_{\mathrm{Newt}}$ for different values of $\Lambda$. The green curve shows the GR energy deposition for comparison.}
\label{Metrica1}
\end{figure}

We can keep $q=0$: In this case, we have three real solutions for $g_{00}=0$ denoting the cosmological horizon ($R_{\mathrm{Ch}}$) along with an outer and inner event horizon. The results for this metric are presented in Fig~\ref{Metrica1} (where for $\dot{Q}$ we have integrated until $R_{Ch}$). The black line, corresponding to $\Lambda=10^{-2}M^{-2}$, shown an enhance respect to GR and it has a different behavior respect to the other curves due to the presence of $R_{\mathrm{Ch}}\sim 16M$.\\
In this parameter range, the model doesn't show significant differences respect to GR.
\subsection*{}
\begin{figure}[h!]
\centering
\includegraphics[scale=0.7]{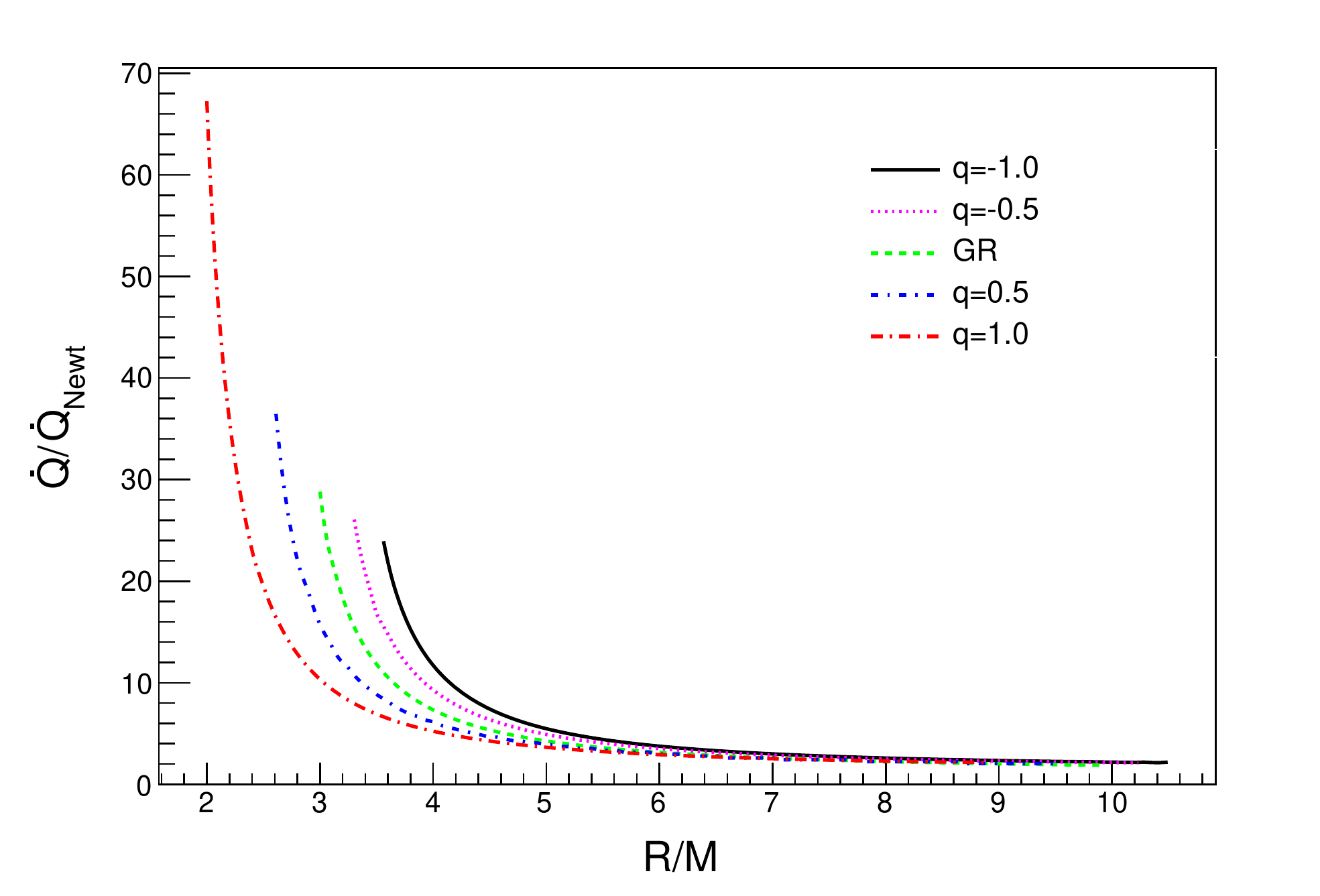}
\caption{Ratio of energy deposition $\dot{Q}$ for metric in Eq.~(\ref{metrica1}) to total Newtonian energy deposition $\dot{Q}_{\mathrm{Newt}}$ for $\Lambda=0$ (similar shape is obtain for different value of $\Lambda$). For different values of $q$ we have different values of the radius of the photosphere. The green curve shows the GR energy deposition for comparison.}
\label{L10-3}
\end{figure}

If we keep $q\neq 0$, we have the shape for the energy deposition described in Fig.~\ref{L10-3}. It is possible to notice the enhancement of the maximum amount of energy deposition, up to a factor $2$, respect to GR case ($q=0$) and the shift in the minimum photosphere radius $R_{ph}$. \\
For values of $R/M$ where energy deposition is not defined in GR, we have however extended the definition $\dot{Q}/\dot{Q}_{\mathrm{Newt}}=\mathcal{F}(M/R)$.
\subsection{Einstein dilaton Gauss Bonnet gravity}
\label{Einstein dilaton Gauss Bonnet gravity}
\begin{figure}[h!]
\centering
\includegraphics[scale=0.7]{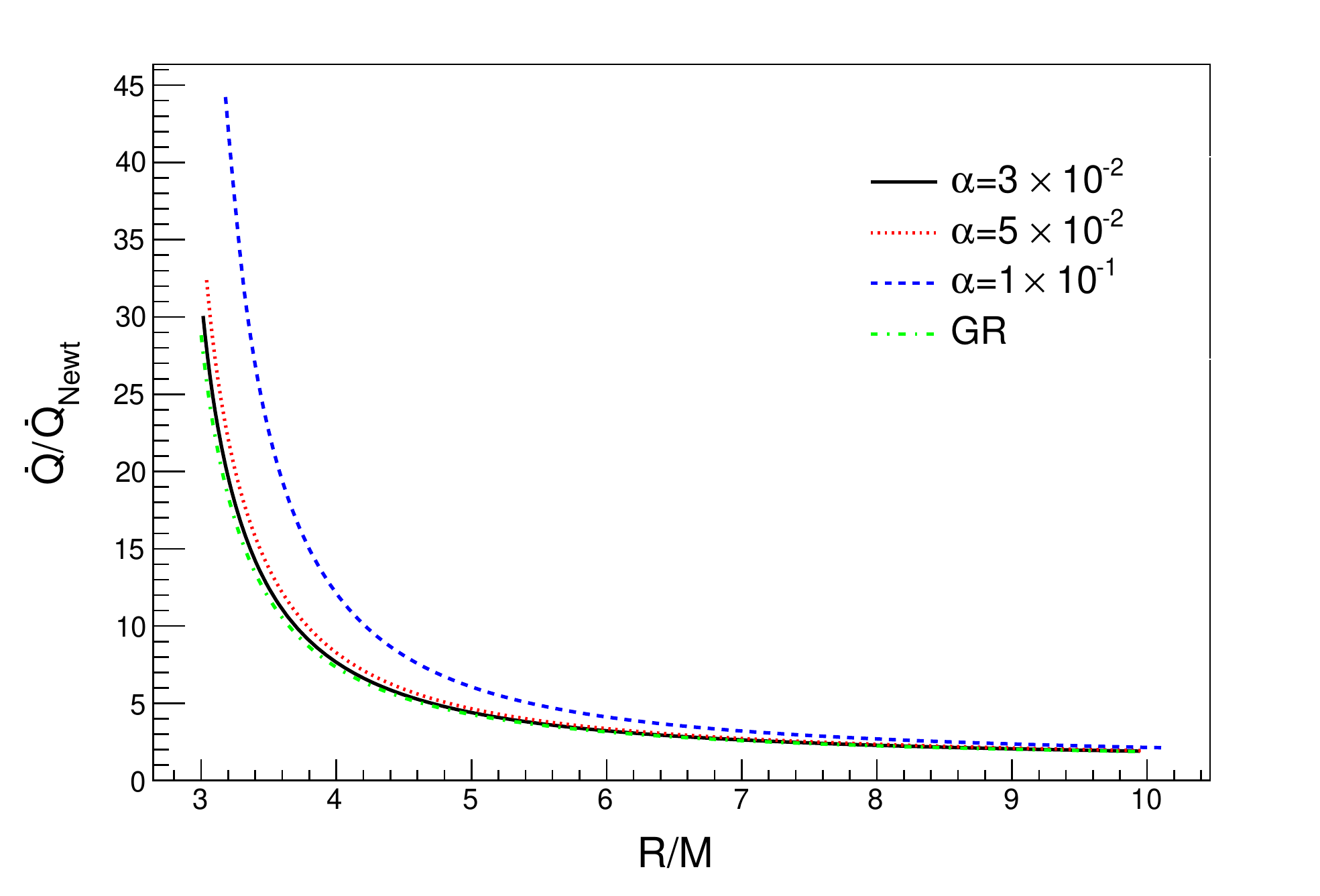}
\caption{Ratio of energy deposition $\dot{Q}$ for metric in Eq.~(\ref{metrica2}) to total Newtonian energy deposition $\dot{Q}_{\mathrm{Newt}}$ for different values of $\bar{\alpha}$. The green curve shows the GR energy deposition for comparison.}
\label{Figmetrica2}
\end{figure}

In this Section we discuss about the solution in spherical symmetry is a subclass of Horndesky theory and correspond to Einstein dilaton Gauss Bonnet gravity. The action is~\citep{Mukherjee:2017fqz}:
\begin{equation}
S=\frac{1}{8\pi}\int d^4x\sqrt{-g}\left(\frac{R}{2}-\frac{1}{2}\partial_{\mu}\psi\partial^{\mu}\psi+\alpha\psi L_{GB}\right) \,\ ,
\end{equation}
where $\psi$ is a scalar field and $L_{GB}$ is the Gauss-Bonnet invariant: $L_{GB}=R^2-4R^{\alpha\beta}R_{\alpha\beta}+R^{\alpha\beta\gamma\delta}R_{\alpha\beta\gamma\delta}$. The solution considered is the Sotiriou-Zhau solution, valid for small $\bar{\alpha}=\alpha/4M^2$ (solution in perturbation theory)~\citep{Sotiriou:2014pfa}:
\begin{equation}
ds^2=-f(r)dt^2+h(r)dr^2+r^2d\Omega^2 \,\ ,
\label{metrica2}
\end{equation}
whit:
\begin{align*}
f(r)&=\left( 1-\frac{2m}{r}\right)\left( 1+\sum_nA_n\bar{a}^n\right) \,\ ;\\
h(r)&=\left( 1-\frac{2m}{r}\right)^{-1}\left( 1+\sum_nB_n\bar{a}^n\right)\,\ .
\end{align*}
where, to the second order:
\begin{align*}
A_1&=B_1=0 \,\ ;\\
A_2&=-\frac{49}{40m^3r}-\frac{49}{20m^2r^2}-\frac{137}{30mr^3}-\frac{7}{15r^4}+\frac{52m}{15r^5}+\frac{40m^2}{3r^6} \,\ ;\\
B_2&=\frac{49}{40m^3r}+\frac{29}{20m^2r^2}+\frac{19}{10mr^3}-\frac{203}{15r^4}-\frac{436m}{15r^5}-\frac{184m^2}{3r^6} \,\ .
\end{align*}
With the above metric we obtain the shape for energy deposition shown in Fig~\ref{Figmetrica2}. The maximum value taken for $\bar{\alpha}$, considering the perturbative regime of the solution, shown an increase of the $50\%$ for the maximum amount of energy deposition respect to GR.

\subsection{Brans Dicke theory}
\label{Brans Dicke theory}
\begin{figure}
\centering
\includegraphics[scale=0.7]{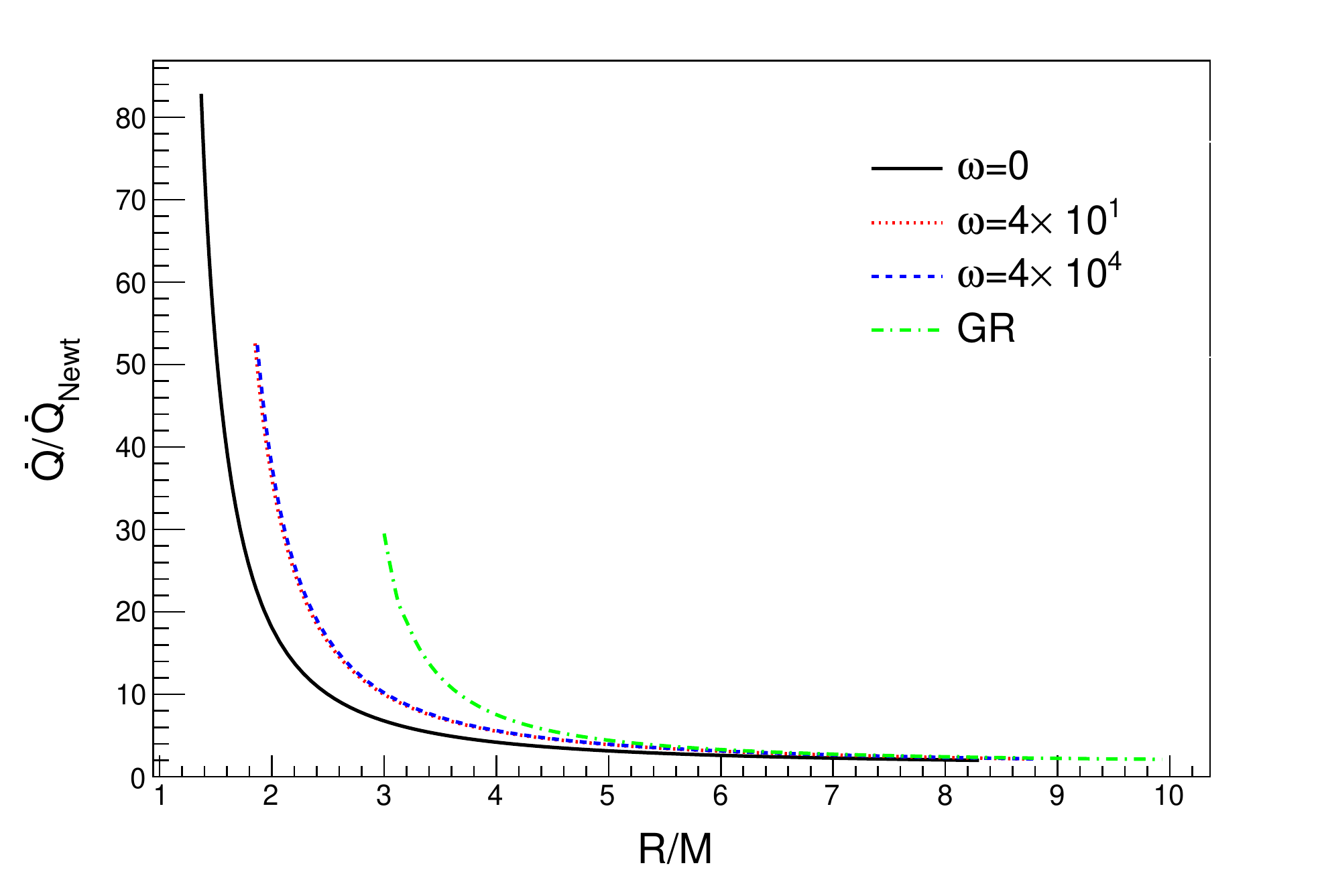}
\caption{Ratio of energy deposition $\dot{Q}$ for metric in Eq.~(\ref{metrica3}) to total Newtonian energy deposition $\dot{Q}_{\mathrm{Newt}}$ for different value of $\omega$. The green curve shows the GR energy deposition for comparison.}
\label{Figmetrica3}
\end{figure}

In this Section we discuss of $\nu\bar{\nu}$ annihilation in the Brans Dicke theory. It represents a generalization of general relativity, where gravitational effects are in part due to geometry, in part due to a scalar field. The action is~\citep{Brans:1961sx}
\begin{equation}
S=\int d^4x\sqrt{-g}\left[\psi R+\frac{16\pi}{c^4}L-\omega(\psi)\right] \,\ ,
 \end{equation} 
 where $L$ is the Lagrangian density of all the matter, including all non-gravitational field, $\psi$ is a scalar field and $\omega$ is its Lagrangian density.\\
 With this Lagrangian, expressing the line element in the isotropic form, we obtain the solution~\citep{Brans:1961sx}
 \begin{equation}
 ds^2=-e^{2\alpha}dt^2+e^{2\beta}\left[dr^2+r^2d\Omega^2\right] \,\ ,
 \label{metrica3}
 \end{equation}
 where
 \begin{align*}
 \lambda&=\sqrt{(C+1)^2-C(1-\frac{C\omega}{2})} \,\ \\
 e^{2\alpha}&=e^{2\alpha_0}\left[\frac{1-\frac{B}{r}}{1+\frac{B}{r}}\right]^{\frac{2}{\lambda}} \,\ ,\\
 e^{2\beta}&=e^{2\beta_0}\left(1+\frac{B}{r}\right)^4\left[\frac{1-\frac{B}{r}}{1+\frac{B}{r}}\right]^{\frac{2(\lambda-C+1)}{\lambda}} \,\ ,\\
 \psi&=\psi_0\left[\frac{1-\frac{B}{r}}{1+\frac{B}{r}}\right]^{-\frac{C}{\lambda}} \,\ ,
 \end{align*}
 with $\omega$ positive constant and
 \begin{align*}
 \alpha_0&=\beta_0=0 \,\ ,\\
 \psi_0&=\frac{4+2\omega}{3+2\omega} \,\ ,\\
 C&\sim-\frac{1}{2+\omega} \,\ ,\\
 B&\sim\frac{M}{2\sqrt{\psi_0}} \,\ .
 \end{align*}
Using this metric, we obtain the shape for energy deposition in Fig.~\ref{Figmetrica3}. Even with this model we have an enhancement of about $50\%$ respect to the maximum value of $\dot{Q}/\dot{Q_{\mathrm{Newt}}}~30$ in GR. 

\subsection{Eddington-inspired Born-Infeld black hole solution}
\label{Eddington-inspired Born-Infeld black hole solution}
\begin{figure}[h!]
    \centering
    \includegraphics[scale=0.7]{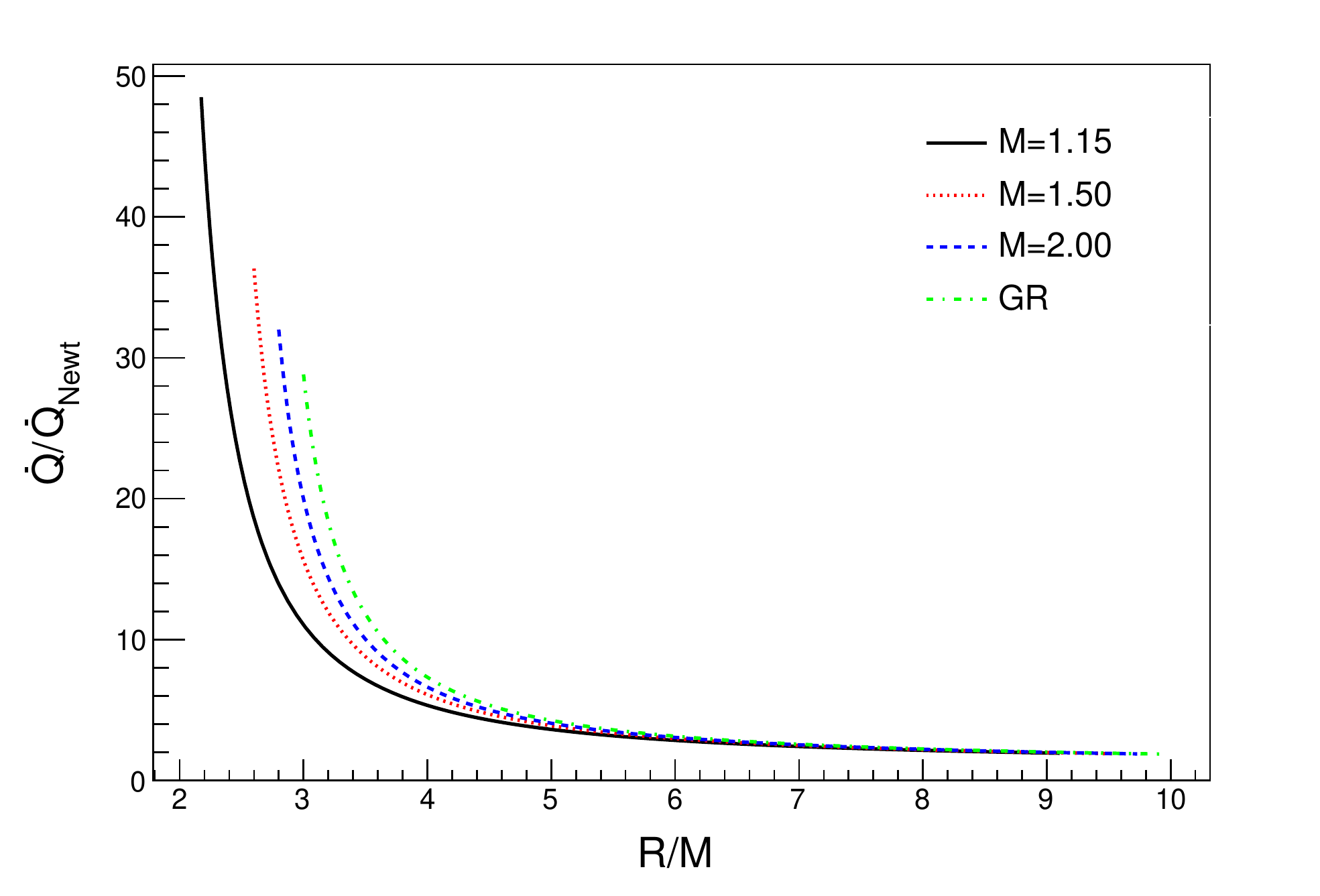}
    \caption{Ratio of total energy deposition $\dot{Q}$ for metric in Eq.~(\ref{metrica4}) to total Newtonian energy deposition $\dot{Q}_{\mathrm{Newt}}$ for $\epsilon=Q=\lambda=1$ and various values of mass. The green curve shows the GR energy deposition for comparison.}
    \label{Metrica_im}
\end{figure}

In this Section we turn our attention on the spherically symmetric solution that Baados and Ferreira considered for the line element in the Born-Infeld model coupled with an electric field~\citep{BeltranJimenez:2017doy}. The action can be written as:
\begin{equation}
    S=\frac{1}{\epsilon k^2}\int d^4x\left[\sqrt{-det\left(g_{\mu\nu}+\epsilon R_{\mu\nu}(\Gamma)\right)}-\lambda\sqrt{-det g_{\mu\nu}}\right]+S_M(g_{\mu\nu},\psi_m) \,\ ,
\end{equation}
where $\psi_m$ is the matter field, $\Gamma$ the connection and $\epsilon$ and $\lambda$ parameters. The line element reads as: 
\begin{equation}
    ds^2=-\psi(r)^2f(r)dt^2+\frac{dr^2}{f(r)}+r^2d\Omega^2 \,\ ,
    \label{metrica4}
\end{equation}
where
\begin{align}
    \psi(r)&=\frac{r^2}{\sqrt{r^4+(\epsilon/\lambda)Q^2}} \,\ ;\\
    \begin{split}
    f(r)&=\frac{r\sqrt{\epsilon Q+\lambda r^4}}{\lambda r^4-\epsilon Q^2}\Bigg[\frac{(3r^2-Q^2-(\lambda-1)r^4/\epsilon)\sqrt{\epsilon Q^2+\lambda r^4}}{3r^3}+\frac{1}{3}\sqrt{\frac{Q^3}{\pi\sqrt{\epsilon\lambda}}}\Gamma^2(\frac{1}{4})\\
    &+\frac{4}{3}\sqrt{\frac{iQ^3}{\sqrt{\epsilon\lambda}}}F\left(i\mathrm{arcsinh}\left(\sqrt{\frac{i}{Q}\sqrt{\frac{\lambda}{\epsilon}}}r\right),-1\right)-2\sqrt{\lambda}M\Bigg] \,\ , 
    \end{split}
\end{align}
with
\begin{equation}
    F(\beta,\alpha)=\int_0^\beta(1-\alpha^2\sin^2\theta)^{-1/2}d\theta \,\ .
\end{equation}
Using this metric we obtain the shape for the energy deposition in Fig.~\ref{Metrica_im}. Even in this case we have a maximum increase of about $50\%$.

\subsection{Born–Infeld generalization of Reissner–Nordstrom solution}
\label{Born–Infeld generalization of Reissner–Nordstrom solution}
\begin{figure}[h!]
    \centering
    \includegraphics[scale=0.7]{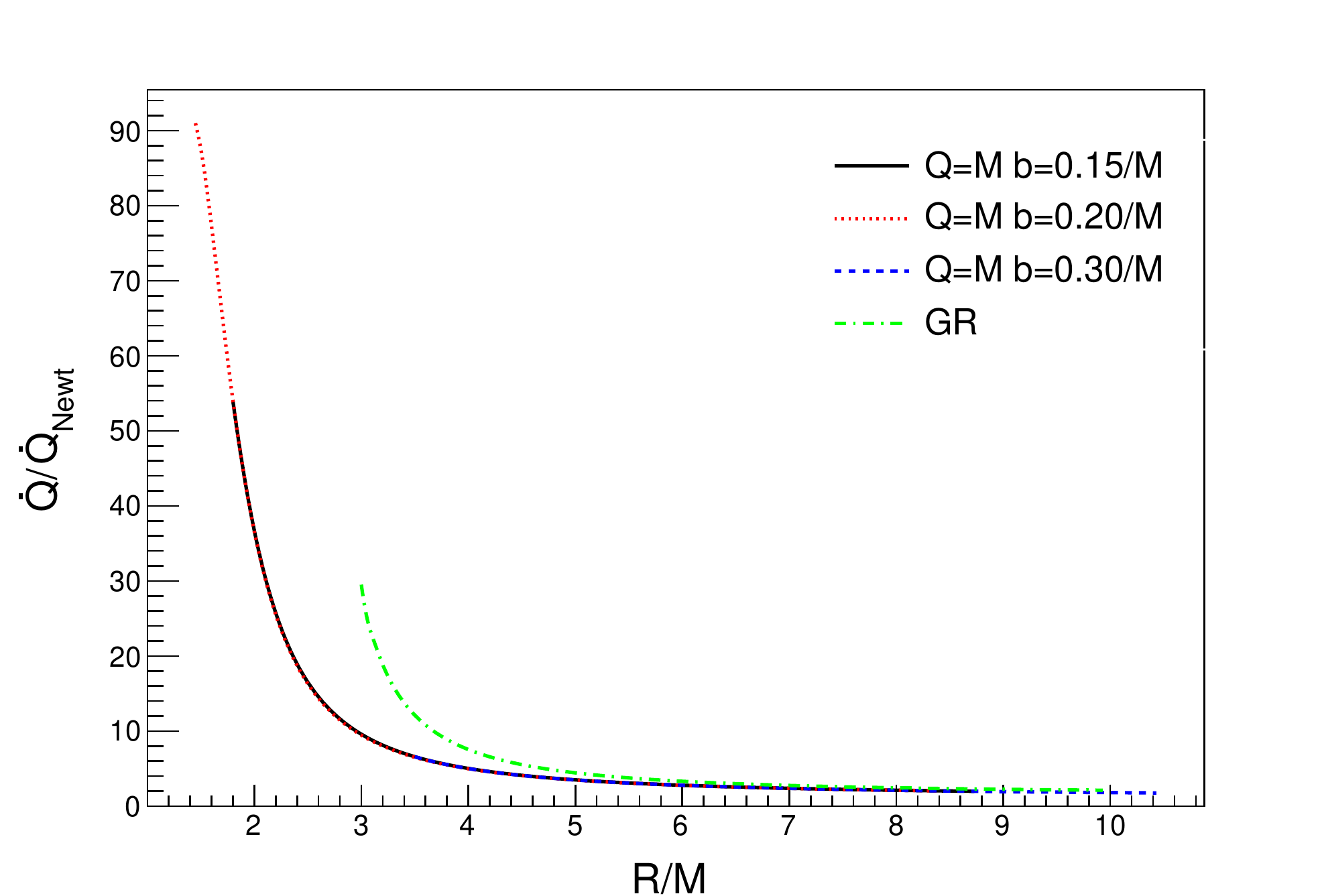}
    \caption{Ratio of total energy deposition $\dot{Q}$ for metric in Eq.~(\ref{Metrica7}) to total Newtonian energy deposition $\dot{Q}_{\mathrm{Newt}}$ for 3 values of the parameter $b$ and $Q=M$. The green curve shows the GR energy deposition for a comparison.}
    \label{Metrica7fig}
\end{figure}

In the Born-Infeld model, it is also possible to write the generalization of Reissner–Nordstrom solution as~\citep{Breton:2002td}:
\begin{equation}
    ds^2=-\psi dt^2+\psi^{-1} dr^2+r^2d\Omega^2 \,\ ,
\end{equation}
where
\begin{equation}
    \psi=1-\frac{2M}{r}+\frac{2}{3}b^2r^2\left( 1-\sqrt{1+\frac{Q^2}{b^2r^4}}\right)+\frac{2Q^2}{3r}\sqrt{\frac{b}{Q}}F\left(arccos\left(\frac{br^2/Q-1}{br^2/Q+1}\right),\frac{1}{\sqrt{2}}\right) \,\ .
    \label{Metrica7}
\end{equation}
The Reissner–Nordstrom solution characterize the final state of a charged star, having as its uncharged limit the Schwarzschild black hole. It is interesting to investigate the neutrino annihilation energy deposition in its nonlinear electromagnetic generalization.\\
With this metric we obtain the shape for the energy deposition in Fig.~\ref{Metrica7fig}. It is possible to see a relevant enhancement (or suppression) of the annihilation energy released, up to $200\%$ in the case $Q=M$ and $b=0.3/M$.

\subsection{Higher derivative gravity analytical solution}
\label{Higher derivative gravity analytical solution}
\begin{figure}[h!]
    \centering
    \includegraphics[scale=0.7]{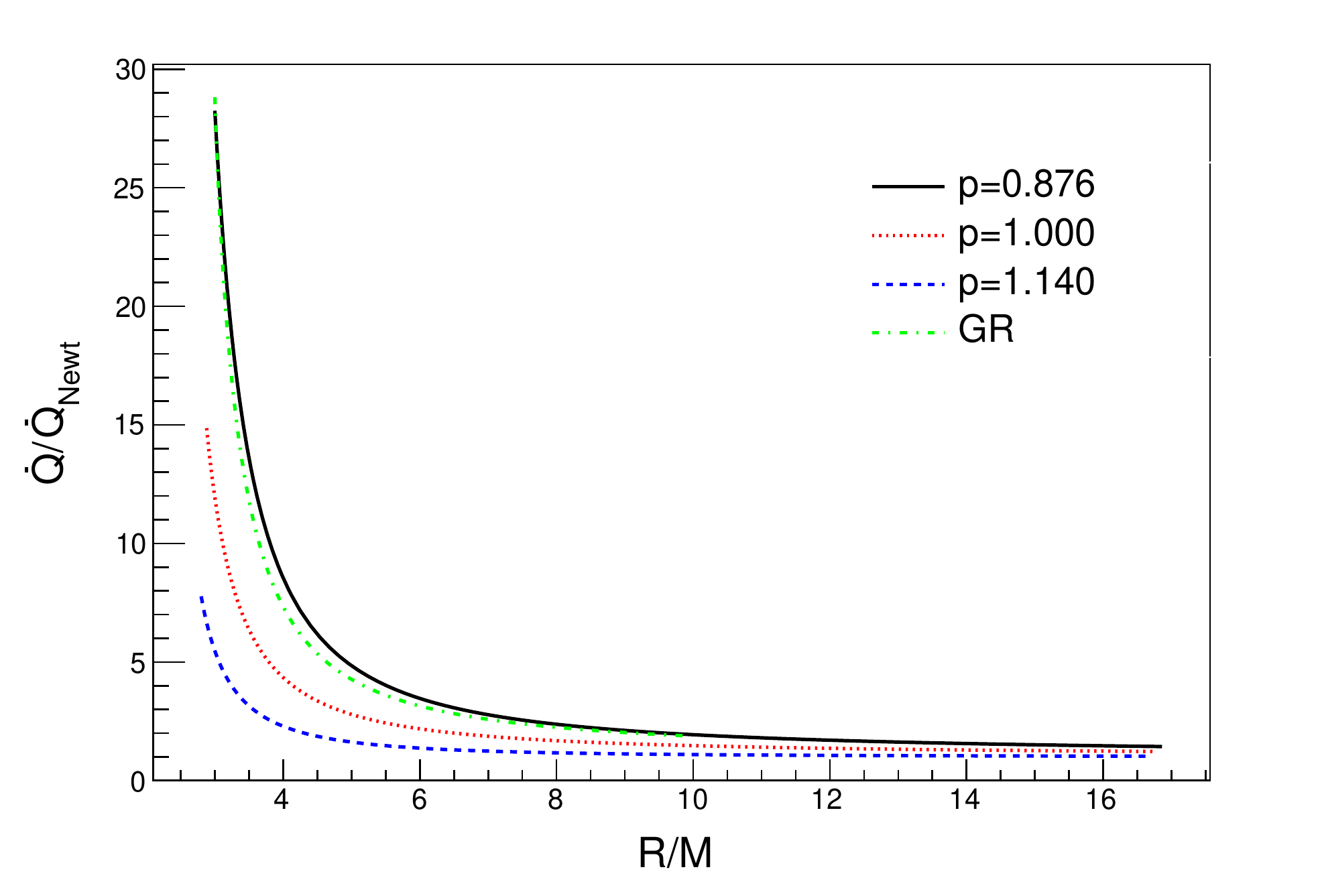}
    \caption{Ratio of total energy deposition $\dot{Q}$ for metric in Eq.~(\ref{metrica5}) to total Newtonian energy deposition $\dot{Q}_{\mathrm{Newt}}$ for tree values of the parameter $p$. The green curve shows the GR energy deposition for comparison. }
    \label{Metrica5}
\end{figure}

Finally,in this Section we study the analytic solution in higher order gravity. The action can be expressed as
\begin{equation}
 S=\int d^4x\sqrt{-g}\left(\gamma R-\alpha C_{\mu\nu\rho\sigma}C^{\mu\nu\rho\sigma}+\beta R^2\right)\,\ ,   
\end{equation}
where $\alpha$, $\beta$ and $\gamma$ are constants and $C^{\mu\nu\rho\sigma}$ is the Weyl tensor. The non-Schwarzschild BH solution is of the form~\citep{Kokkotas:2017zwt}:
\begin{equation}
    ds^2=-\left(1-\frac{r_0}{r}\right)A(r,p)dt^2+\frac{B(r,p)^2dr^2}{\left(1-\frac{r_0}{r}\right)A(r,p)}+r^2d\Omega^2 \,\ .
    \label{metrica5}
\end{equation}
The expressions of the metric components $A(r,p)$ and $B(r,p)$ are quite involved and are not reported here  (see Ref. ~\citep{Kokkotas:2017zwt}), while $p$ is a parameters of the model defined as
\begin{equation}
    p=\frac{r_0}{2\alpha} \,\ ,
\end{equation}
where $r_0$ is the horizon radius. The parameter $p$ goes from $0.876$, corresponding to the merger of Schwarzschild and non-Schwarzschild solution, to $1.14$ where the non-Schwarzschild solution almost vanishes
\footnote{More precisely,  the Schwarzschild metric is also an exact solution of the Einstein-Weyl theory for all $p$. However, at some minimal non-zero value of $p$, $p_{min}$, there appears (in addition to
the Schwarzschild solution) a non-Schwarzschild branch, that is a solution that 
describes the asymptotically flat black hole, characterized by a mass which decreases as $p$ grows, and vanishes at some $p_{max}$. The range of values of $p$ are ones discussed in the text, i.e. $p\in [0,84; 1.14]$.}.
The behaviour of the energy deposition, for high values of the parameter $p$, is represented in Fig.~\ref{Metrica5}. In such a case, we obtain a reduction of the energy released by the neutrino pair annihilation, up to a reduction of a factor $6$ for $p=1.14$.

\section{Gamma Ray Burst analysis in extended theories of gravity}
\label{Gamma Ray Burst analysis}

Results of the previous Section show that modified gravity provides, in almost all cases, a non-trivial deviation from General Relativity behaviour. Such a deviation is extremely relevant for neutron star (NS) and black hole evolution, as well as for the Gamma-Ray Burst (GRB) phenomena. In the latter case, to which we are interested in this Section, we shall highlight that neutrino pairs annihilation might partially contribute.\\    
GRBs and their possible connection with neutrino production in compact stars is a field of high current interest. They represent probably the biggest mystery in high energy astronomy. GRBs are of two kinds~\citep{2}: 
\begin{itemize}
    \item [1.] Long duration bursts characterized by a duration is in the range [$2~\mathrm{s}$, several minutes] with the average of $\sim O(30~\mathrm{s})$;
    \item [2.] Short duration bursts characterized by a duration in the range [$O(10^{-3}~\mathrm{s})$, $2~\mathrm{s}$] with the average of $\sim 0.3~\mathrm{s}$.
\end{itemize}
First evidence of long GRBs associated to SNe were derived by studying GRB 980425 \citep{3} and GRB 030329 \citep{4,41,42}. Regarding short GRBs, there is a scarcity of information. 
Particularly relevant in these frameworks is the reaction $\nu+{\bar \nu}\to e^-+ e^+$, since the $e^{\pm}$ pairs may further give rise to gamma rays which could be a possible explanation of the observed GRBs. Previous analysis of the reaction $\nu+{\bar \nu}\to e^-+ e^+$ near a neutron star based on Newtonian gravity (i.e. in the regime $r\gg R_s$, where $R_s$ is the Schwarzschild radius) has been performed in Refs. \citep{5,6}. The inclusion of gravitational effects for static stars was developed in \citep{Salmonson:1999es, 8}. The inclusion of rotation of stars was studied in \citep{Bhattacharyya:2009nm,kovacs}. These effects are non negligible in strong gravitational regimes, as well as in theories of gravity that extended GR, as we shall discuss. Indeed, it is possible to make a simple evaluation of the energy emitted from neutrino pair annihilation considering Eq.~(\ref{contoQ}), with a total luminosity from NS for neutrinos of the order $\sim O(10^{53}~\mathrm{erg/s})$ and radius $R$ of the the order  $\sim O(20~\mathrm{km})$~\citep{Perego:2017fho}:
\begin{align}
 \dot{Q}&=1.2116\times 1.09\times 10^{-5}(10^2)^{9/4}2^{-3/2}\mathrm{\frac{erg}{s}}\mathcal{F}(R) \\&=1.48\times10^{50}~\mathrm{\frac{erg}{s}}\mathcal{F}(R) \,\ ,   
\end{align}
where we use the smallest value possible for $D$ and the function $\mathcal{F}(R)$ depends on the theory under consideration (see Eq. (\ref{contoQ})). For GR, $\mathcal{F}\sim 30$ considering that in a neutron star neutrinos are emitted from the photosphere. Therefore, we obtain $\dot{Q}\sim 4.41\times 10^{51}~\mathrm{erg/s}$ which is inferior to the GRB emitted energy rate $\sim 5\times 10^{52}\mathrm{erg/s}$~\citep{Bhattacharyya:2009nm}.
As shown in our work, the maximum of $\mathcal{F}$ in theories beyond GR can be different and, in some cases, we have a variation up to a factor $3$, obtaining
\begin{equation}
    \dot{Q}_{\mathrm{max}}=1.3\times 10^{52}~\mathrm{\frac{erg}{s}} \,,
\end{equation}
which is almost of the same order of magnitude of the GRB emission energy rate.

As these results suggest, it is possible that theories of gravity beyond GR may efficiently contribute, as a consequence of the reduction of the photosphere radius (hence the gravitational effects turn out to be enhanced), to the GRBs emission through the neutrino pair annihilation mechanism. This could be a test-bed for probing non-standard gravity described by GR. 


\section{Conclusions}
\label{Conclusions}

In this paper, we have provided a systematic way to treat the neutrino pair annihilation $\nu\bar{\nu}\rightarrow e^+e^-$ and we have analyzed the process in the framework of various models of gravity beyond General Relativity. We extended the general relativistic calculations of neutrino heating rates, shown an increase of a factor $\sim 27$ respect to Newtonian calculations, to Charged  Galileon,  Einstein dilaton  Gauss-Bonnet,  Brans  Dicke,  Eddington-inspired  Born-Infeld, Born–Infeld generalization of Reissner–Nordstrom solution and Higher derivative gravity model respectively finding some relevant differences in the minimum photon sphere radius $R_{ph}$ and in the maximum in the energy deposition rate near to $R_{ph}$.

We have shown that in the opportune range of parameters, one can obtain relevant enhancement (up to a factor $3$) or suppression (up to a factor $1/3$) of the maximum rate  and a shift in the minimum photosphere radius.

These results can be extremely important in the context of neutron star merging (for which one has to consider the value of the rate at $R_{ph}$) or in the supernovae, where it is relevant to consider $R=4$-$5~M$.
In particular, we would like to state that the enhancement shown in some model of gravity beyond GR could be relevant for GRBs, for which neutrino pairs annihilation has been proposed as a possible source. Indeed, although the source of gamma-ray bursts is still undetermined, neutrino pairs annihilation is not the best candidate due to the fact that some extra energy is needed. More precisely, in the case of short GRBs, a factor $2~\mathrm{or}~3$ up to $10$~\citep{Perego:2017fho} is needed to generate all the gamma-ray bursts observed. As we have highlighted in this paper, in some modified theories of General Relativity it is possible to partially obtain the extra energy needed. Moreover, it is important to remark that in our analysis we have not considered any trapping for the neutrinos, but they may be trapped producing an increase in the temperature ~\citep{Ghosh:1995dn}. Thus other changing in the energy deposition rate could be obtained.

We finally conclude pointing out that deviations from General Relativity could increase significantly all the energy deposition processes in NS and Supernova envelope, suggesting that further investigations have to be done in these frameworks.

\section*{Acknowledgements}
The work of G.L. and L.M. is supported by the Italian Istituto Nazionale di Fisica Nucleare (INFN) through the ``QGSKY'' project and by Ministero dell'Istruzione, Universit\`a e Ricerca (MIUR).\\
The computational work has been executed on the IT resources of the ReCaS-Bari data center, which have been made available by two projects financed by the MIUR (Italian Ministry for Education, University and Re-search) in the "PON Ricerca e Competitività 2007-2013" Program: ReCaS (Azione I - Interventi di rafforzamento strutturale, PONa3\_00052, Avviso 254/Ric) and PRISMA (Asse II - Sostegno all'innovazione, PON04a2\_A)

\bibliography{sample63}{}
\bibliographystyle{aasjournal} 

\end{document}